\hfuzz 2pt
\font\titlefont=cmbx10 scaled\magstep1
\magnification=\magstep1

\null
\vskip 1.5cm
\centerline{\titlefont ON BELL'S LOCALITY TESTS}
\medskip
\centerline{\titlefont WITH NEUTRAL KAONS}
\vskip 3.5cm
\centerline{\bf F. Benatti}
\smallskip
\centerline{Dipartimento di Fisica Teorica, Universit\`a di Trieste}
\centerline{Strada Costiera 11, 34014 Trieste, Italy}
\centerline{and}
\centerline{Istituto Nazionale di Fisica Nucleare, Sezione di 
Trieste}
\vskip 1cm
\centerline{\bf R. Floreanini}
\smallskip
\centerline{Istituto Nazionale di Fisica Nucleare, Sezione di 
Trieste}
\centerline{Dipartimento di Fisica Teorica, Universit\`a di Trieste}
\centerline{Strada Costiera 11, 34014 Trieste, Italy}
\vskip 3cm
\centerline{\bf Abstract}
\smallskip
\midinsert
\narrower\narrower\noindent
We present some general considerations on possible experimental
tests of Bell's locality in the neutral kaon system, and
comment on the recent literature on the topic.
\endinsert
\bigskip
\vfil\eject

Recently, new reports discussing possible tests of the condition of
Bell's locality have appeared.[1, 2] They propose a direct experimental
check of a ``new'' Bell-like inequality for the system of
correlated neutral kaons coming from the decay of a $\phi$-meson.

In this note, we would like to present a few basic
considerations concerning some aspects of possible
tests of local deterministic extensions
of quantum mechanics using the $K^0$-$\overline{K^0}$ system,
that seem to have been overlooked in the works 
quoted above. In particular, we want to emphasize that:
{\it i)} the proposed ``new''
Bell's inequality has already been discussed before,
although in slightly different form;[3, 4]\ {\it ii)}
no direct experimental
test of this inequality is presently possible at available
experimental setups, while actual data on a different test of Bell's
locality proposed in [5] will be available soon.
We stress that the Bell inequality discussed in [5] is
completely general within the stated assumption of stochastic
independence of correlated kaons decays.
\bigskip

The two neutral kaons that originate from the decay of a $\phi$-meson,
although flying apart with opposite momentum (in the $\phi$ rest frame),
remain nevertheless quantum mechanically correlated in a way that is
very similar to the entanglement of two spin-1/2 particles in a
singlet state. This allows performing fundamental
tests on the behaviour of entangled systems 
by studying the time-evolution of
certain observables of the two kaons 
(for a discussion, see [3, 6, 7] and references therein).

Typical observables that can be studied are double
decay probabilities ${\cal P}(f_1,\tau_1; f_2,\tau_2)$,
{\it i.e.} the probabilities that one kaon decays
into a final state $f_1$ at proper time $\tau_1$, while the other kaon 
decays into the final state $f_2$ at proper time $\tau_2$. 
The requirement of the condition of Bell's locality (see later)
results into certain inequalities that the probabilities $\cal P$
should satisfy (Bell's inequalities). A general introduction
to these topics can be found in Refs.[8-11].

Notice that these decay probabilities ${\cal P}(f_1,\tau_1; f_2,\tau_2)$
obviously  differ from the more often considered decay rates
$\Gamma(f_1,\tau_1; f_2,\tau_2)$.[12, 13] Nevertheless, they can easily be
expressed in terms of the latter, as clearly discussed
in the second reference of [5] (see also \hbox{[14, 15]}).
In this respect, the probabilities ${\cal P}(f_1,\tau_1; f_2,\tau_2)$
are perfectly measurable quantities, related to the
actual experimental counting of the partial occurance 
of the various final decay states for the two kaons.
Finally, the overall normalization of the probabilities $\cal P$
({\it i.e.} a factor taking care of the absolute branching 
ratio for the various decay channels) is irrelevant for the 
considerations leading to the derivation of the Bell's inequalities
(for details, see \hbox{[9, 5]}).

In deriving these inequalities, one assumes that the system of the
two neutral kaons can be fully described by a set of variables 
$\{\lambda\}$. Then, for the average probability 
${\cal P}(f_1,\tau_1; f_2,\tau_2)$ one explicitly has:
$$
{\cal P}(f_1,\tau_1;f_2,\tau_2)=\int d\lambda\, \rho(\lambda)\,
p_\lambda(f_1,\tau_1;f_2,\tau_2)\ ,\eqno(1)
$$
where $\rho(\lambda)$ is a normalized probability density characterizing
the ensemble of initial $\phi$ particles and, for fixed $\{\lambda\}$,
$p_\lambda(f_1,\tau_1;f_2,\tau_2)$ is a well-defined probability
of detecting a final decay state $(f_1,\tau_1)$ for one of the two kaons
and a final decay state $(f_2,\tau_2)$ for the second one. 
It is important to observe that, since no assumption have been
made on the set $\{\lambda\}$, this description is very general,
and surely can be made to agree with the results of ordinary
quantum mechanics.

On the contrary, the hypothesis of Bell's locality adds a further
condition. It is based on the observation that
the decays of the two kaons coming from a $\phi$-meson are localized
events, usually very well spatially separated; one is then led
to consider a description where the joint
probability $p_\lambda(f_1,\tau_1;f_2,\tau_2)$ is naturally
expressed in a product form:
$$
p_\lambda(f_1,\tau_1;f_2,\tau_2)=p_\lambda(f_1,\tau_1;-,\tau_2)\ 
p_\lambda(-,\tau_1;f_2,\tau_2)\ ,\eqno(2)
$$
where $p_\lambda(f_1,\tau_1;-,\tau_2)$ is the probability density of
detecting a final state $(f_1,\tau_1)$ for one of the kaons
and any state $(-,\tau_2)$ for the second one, and similarly
for $p_\lambda(-,\tau_1;f_2,\tau_2)$.
The condition of Bell's locality in (2) leads to measurable consequences
that can be experimentally tested.

In particular, by further assuming that the decay of one
kaon is stochastically independent from the decay of the second one,
from (2) one easily derives the following inequality:
$$
{\cal P}(f_3,\tau;f_2,\tau)-{\cal P}(f_3,\tau; f_1,\tau)\leq
{\cal P}(f_1,\tau;f_2,\tau)\ .\eqno(3)
$$
As stressed in [5], this inequality is not satisfied by all theories
describing the two kaons in terms of the variables $\{\lambda\}$;
indeed, in deriving (3) from (2), one (implicitly) needs to assume
the absence of variables $\lambda$ that, without violating Bell's locality (2), 
might nevertheless correlate the two decays.
This further assumption is however unavoidable if one uses
the standard effective two-dimensional theory to model the 
$K^0$-$\overline{K^0}$ system:\hbox{[16, 13]} this choice
is the only one compatible with the description of the final decay states 
as actual asymptotic particles,
and therefore directly accessible to the experiment.

The final decay states in (3) are totally arbitrary and can be chosen
at will. (This freedom directly corresponds to the choice of the
photon polarizations in conventional Bell's inequality tests.[8-11])
When $f_1$, $f_2$ are taken to be two-pion states and $f_3$
a semileptonic state, from (3) one deduces [5]
$$
\Big|{\cal P}(\pi^-\ell^+\nu,\tau;2\pi^0,\tau)
-{\cal P}(\pi^-\ell^+\nu,\tau;\pi^+\pi^-,\tau)\Big|\leq
{\cal P}(2\pi^0,\tau;\pi^+\pi^-,\tau)\ .\eqno(4)
$$
A similar relation holds when the semileptonic state $\pi^-\ell^+\nu$
is substituted with $\pi^+\ell^-\bar\nu$.

The various joint probabilities in (4) can be computed using standard
particle physics techniques and the usual effective phenomenological
description of the neutral kaon system, allowing $CP$ as well as
$CPT$ and $\Delta S=\Delta Q$ violating effects.
It turns out that the particular choice of the final states above
allows to express (4) only in terms of the parameter $\varepsilon'$,
that describes direct $CP$ violating effects in the $2\pi$ final states:
$$
\big|{\cal R}e(\varepsilon')\big|\leq 3\, |\varepsilon'|^2\ .\eqno(5)
$$
In practice, the assumption of Bell's locality (2) for the system
of correlated neutral kaons coming from the decay of a $\phi$ predicts
that $\varepsilon'$ must satisfy the condition (5).
It should be stressed that ignoring $CP$ violations in decay processes,
as done in all discussions on Bell's inequalities in the 
$K^0$-$\overline{K^0}$ system, would reduce (5) to a trivial identity.
The parameter $\varepsilon'$ is accessible to the experiment, which makes 
even more transparent that the inequality (4) involve measurable quantities.
Any experimental determination of this parameter, in any actual setup,
not necessarily involving correlated kaons, would automatically be
a check of Bell's locality, within the assumption of stochastic independence
of kaon decays.

Let us point out that (5) is not a test on the 
theoretical framework in which the parameter $\varepsilon'$ can
be estimated.[13, 17] As any test,
also the ones based on the inequality (5) are significant only 
when they give a negative result, 
saying that Bell's inequality (4) is violated.
Only in this case, one is able to experimentally exclude
a large set of local deterministic extensions of quantum mechanics.
If the experimental data turn out to be compatible with
a vanishing value for $\varepsilon'$ (result predicted by the
so-called superweak phenomenological model [18]), 
than simply the inequality (5)
would not provide any criterion to exclude those extensions of
quantum mechanics, and one would need to look for different tests.
However, if the experiments
\footnote{$\!\!\!^\dagger$}{
The results of the fixed target experiments KTEV at Fermilab and NA48 at CERN 
on the measure of ${\cal R}e(\varepsilon'/\varepsilon)$
are expected soon, while
the KLOE apparatus at the DA$\Phi$NE $\phi$-factory in Frascati
will start collecting data in the near future.}
confirm the theoretical estimates, which roughly predict a value
of order $10^{-6}$ for the modulus of $\varepsilon'$,
with a phase close to $\pi/4$,
then the condition (5) will be violated by several orders of magnitude,
resulting in one of the best tests of Bell's locality.

The discussion that led the authors of [1, 2] to the derivation of their
``new'' Bell's inequality follows very closely the considerations
presented so far. Indeed, for their final analysis they use the
very same inequality in (3), although the final decay states
$f_1$, $f_2$ and $f_3$ are now replaced by suitable states,
obtained by making the kaons interact with a slab of material;
further, in their analysis $CP$ conservation is assumed.
This idea of using the so-called regeneration phenomena for 
a direct experimental test of
Bell's inequalities in a $\phi$-factory
has been already considered and analyzed
in detail in [3, 4]. There, a test of Bell's locality was reduced
to a measure of the real part of the regeneration parameter $\eta$;
a very similar idea is proposed as ``new'' in [1, 2].

Further, as pointed out also in [3, 4], it is technically very
difficult to place a regeneration slab close 
enough to the $e^+$-$\, e^-$ beam in order to actually
perform the proposed Bell's locality test; the only practical way out
is to use the walls of the actual beam-pipe of the accelerator
as a regeneration slab: this is indeed sufficently close to the
$\phi$-decay vertex to allow proper identification of the
$K_S$ mesons, before their decay. 
However, this possibility would put some restrictions
on the class of possible local-deterministic extensions of
quantum mechanics that can be excluded by that experimental
setup: in particular, one is again forced to assume
the stochastic independence of kaon decays.

In conclusion, for the forseeable future the only direct test 
of Bell's inequality in the neutral kaon system that can be actually performed  
remains the measure of the phenomenological parameter $\varepsilon'$.
A non-vanishing value for this parameter in agreement with the
theoretical predictions would allow the exclusion of a wide class
of local deterministic extensions of quantum mechanics
with an accuracy never reached in the past in any direct experimental
tests of Bell's locality.

\vfill\eject

\centerline{\bf REFERENCES}
\bigskip

\item{1.} B. Ancochea, A. Bramon and M. Nowakowski,
Bell-inequalities for $K^0\overline{K^0}$ pairs from $\phi$-resonance
decays, {\tt hep-ph/9811404}
\smallskip
\item{2.} A. Bramon and M. Nowakowski, Bell-inequalities for entangled
pairs of neutral kaons, {\tt hep-ph/9811406}
\smallskip
\item{3.} A. Di Domenico, Nucl. Phys. {\bf B450} (1995) 293
\smallskip
\item{4.} A. Di Domenico, in {\it Workshop on K Physics},
L. Iconomidou-Fayard, ed., (Editions Frontieres, Gif-sur-Yvette (France),
1997)
\smallskip
\item{5.} F. Benatti and R. Floreanini, Phys. Rev. D {\bf 57} (1998) 1332
and {\tt hep-ph/9712274}
\smallskip
\item{6.} G. Ghirardi, R. Grassi and T. Weber, in the Proceedings
of the {\it Workshop on Physics and Detectors for DA${\mit\Phi}$NE},
G. Pancheri, ed., (INFN-LNF, Frascati, 1991)
\smallskip
\item{7.} P.H. Eberhard, Nucl. Phys. {\bf B398} (1993) 155
\smallskip
\item{8.} J. Bell, {\it Speakable and Unspeakable in Quantum Mechanics},
(Cambridge University Press, Cambridge, 1987)
\smallskip
\item{9.} J.F. Clauser and M.A. Horne, Phys. Rev. D {\bf 10} (1974) 526
\smallskip
\item{10.} J.F. Clauser and A. Shimony, Rep. Prog. Phys. {\bf 41} (1978) 1881
\smallskip
\item{11.} M. Redhead, {\it Incompleteness, Nonlocality and Realism},
(Clarendon Press, Oxford, 1987)
\smallskip
\item{12.} See: C.D. Buchanan, R. Cousins, C. Dib, R.D. Peccei and
J. Quackenbush, Phys. Rev. D {\bf 45} (1992) 4088, and references therein
\smallskip
\item{13.} L. Maiani, in {\it The Second Da$\,\mit\Phi$ne Physics Handbook}, 
L. Maiani, G. Pancheri and N. Paver, eds., (INFN, Frascati, 1995)
\smallskip
\item{14.} F. Benatti and R. Floreanini, Phys. Lett. {\bf B 401} (1997) 337
\smallskip
\item{15.} F. Benatti and R. Floreanini, Nucl. Phys. {\bf B 511} (1998) 550
\smallskip
\item{16.} T.D. Lee and C.S. Wu, Ann. Rev. Nucl. Sci. {\bf 16} (1966) 511
\smallskip
\item{17.} For a recent review, see: S. Bertolini, M. Fabbrichesi and J.O. Eeg,
Estimating $\varepsilon'/\varepsilon$, {\tt hep-ph/9802405}
\smallskip
\item{18.} L. Wolfenstein, Phys. Rev. Lett. {\bf 13} (1964) 562

\bye